\documentclass[a4paper,12pt]{article}
\usepackage{epsfig}
\usepackage{amssymb}
\usepackage{amsfonts}
\usepackage{graphics}
\usepackage{color,amsmath,bm}

\newskip\humongous \humongous=0pt plus 1000pt minus 1000pt

\newif\ifdtup

\catcode`\@=11

\@addtoreset{equation}{section}

\def\@normalsize{\@setsize\normalsize{15pt}\xiipt\@xiipt
\abovedisplayskip 14pt plus3pt minus3pt%
\belowdisplayskip \abovedisplayskip
\abovedisplayshortskip \z@ plus3pt%
\belowdisplayshortskip 7pt plus3.5pt minus0pt}

\def\small{\@setsize\small{13.6pt}\xipt\@xipt
\abovedisplayskip 13pt plus3pt minus3pt%
\belowdisplayskip \abovedisplayskip
\abovedisplayshortskip \z@ plus3pt%
\belowdisplayshortskip 7pt plus3.5pt minus0pt
\def\@listi{\parsep 4.5pt plus 2pt minus 1pt
     \itemsep \parsep
     \topsep 9pt plus 3pt minus 3pt}}

\relax

\catcode`@=12

\catcode`\@=11

\def\section{\@startsection{section}{1}{\z@}{3.5ex plus 1ex minus
   .2ex}{2.3ex plus .2ex}{\large\bf}}

\def\thesection{\arabic{section}}

\def\appendix{\setcounter{section}{0}
\def\thesection{\Alph{section}}}

\begin{document}

\newcommand{\beq}{\begin{equation}}
\newcommand{\eeq}{\end{equation}}
\newcommand{\bea}{\begin{eqnarray}}
\newcommand{\eea}{\end{eqnarray}}

\def\de{\partial}
\def\eps{\varepsilon}
\def\V{{\sf V}}
\def\d{d}

\begin{titlepage}

\renewcommand{\thefootnote}{\fnsymbol{footnote}}
\bigskip

\bigskip
\bigskip
\bigskip
\bigskip

\begin{center}
{\Large  {\bf  Multi-vortices are Wall Vortices: \\ A Numerical
Proof}
 }
\end{center}

\renewcommand{\thefootnote}{\fnsymbol{footnote}}
\bigskip
\begin{center}
{\large   Stefano {\sc Bolognesi}\footnote{bolognesi@nbi.dk}}
 and
 {\large Sven
Bjarke {\sc Gudnason}\footnote{gudnason@nbi.dk}}
 \vskip 0.20cm
\end{center}

\begin{center}
{\it      \footnotesize

The Niels Bohr Institute, Blegdamsvej 17, DK-2100 Copenhagen \O, Denmark} \\
\end {center}

\renewcommand{\thefootnote}{\arabic{footnote}}

\setcounter{footnote}{0}

\bigskip
\bigskip
\bigskip

\noindent
\begin{center} {\bf Abstract} \end{center}

We study the Abrikosov-Nielsen-Olesen multi-vortices. Using a
numerical code we are able to solve the vortex equations with
winding number up to $n=25,000$. We can thus check the wall vortex
conjecture previously made in \cite{wallandfluxes,Bolognesi:2005ty}.
The numerical results show a remarkable agreement with the
theoretical predictions.

\vfill

\begin{flushleft}
December, 2005
\end{flushleft}

\end{titlepage}

\bigskip

\hfill{}

\tableofcontents

\section{Introduction} \label{uno}

The idea of the wall vortex emerged in \cite{wallandfluxes} when we
faced the following situation. Consider a theory that has two
degenerate vacua, one in the Higgs phase and another in the Coulomb
phase. Then take a domain wall that interpolates between the two
vacua and place a monopole on the Coulomb side. What happens if we
continuously move the monopole towards the Higgs phase? It is known
that a monopole in the Higgs phase cannot exist by itself and must
be confined. The only reasonable thing that can happen is that the
monopole wall system is continuously transformed into a
monopole-vortex-wall. This implies that the confining vortex in the
Higgs phase can be continuously transformed into a wall or, in
another way, is made of the same stuff as the wall. Thus we call it
a wall vortex. Now it is easy to understand what are the forces that
keep the vortex soliton together. The wall tension that tends to
squeeze the vortex, is nothing but the Derrick \cite{Derrick:1964ww}
collapse force coming from the scalar part of the action $\de \phi
\de \phi + V(\phi)$. What prevents the soliton from collapsing is a
pressure coming from the magnetic flux inside the vortex. The
magnetic energy is  $B^2 R^2$, but we should keep in mind that when
varying the radius of the cylinder, what remains constant is not the
magnetic field but the flux. In terms of the flux, the magnetic
energy is $\Phi_B / R^2$ and thus a pressure term.

The wall vortex argument, up to this point, was only a qualitative
way to understand the continuous transition from a wall to a vortex
and also the forces that bind together the soliton. In order to
apply this idea also quantitatively we need to find a regime in
which the radius of the vortex $R_V$ becomes much greater than the
thickness of the domain wall $\Delta_W$. We find that this is
exactly the large $n$ limit where $n$ is the number of quanta of
magnetic fluxes carried by the vortex. It is easy to understand why
this happens. The domain wall does not know anything about the
magnetic flux and thus its thickness is independent of $n$. On the
contrary the radius of the vortex depends on $n$, in particular it
grows as $n$ grows. This means that the ratio $\Delta_W/R_V$ can be
made arbitrarily small by increasing the parameter $n$.

In \cite{Bolognesi:2005ty} we applied the wall vortex argument to
the general Abelian-Higgs model. In this case there is only one true
vacuum, the Higgs vacuum. The Coulomb phase is not a true vacuum of
the theory, but due to the symmetry of the problem, it is always a
stationary point of the potential. Is it possible, that also in this
case, in the large $n$ limit, the vortex becomes a wall vortex? This
was the conjecture made in \cite{Bolognesi:2005ty}. The aim of this
paper is to prove this conjecture by means of numerical computation.

We organize the paper in the following way. In Section \ref{theory}
we give a review of the wall vortex idea and we state the
theoretical arguments that support the conjecture. Then in Section
\ref{numerical} we present the numerical proof of the conjecture. In
Section \ref{degenerate} we analyze the original problem when the
Coulomb and Higgs phases are degenerate. Finally in Section
\ref{discussion} we conclude with a discussion about the relation
between different potentials in the large $n$ limit.

\section{Theoretical Section} \label{theory}

In this section we provide the theoretical analysis that leads to the
wall vortex conjecture. A lot of independent arguments support the
wall vortex limit. Part of these arguments are new and part are just
a review of \cite{wallandfluxes,Bolognesi:2005ty}.

The theory under consideration is the Abelian-Higgs model, which is
the relativistic version of the Ginzburg-Landau theory of
superconductivity \cite{Ginzburg:1950sr}. It is a $U(1)$ gauge
theory coupled to a charged scalar field $q$
\begin{equation}
\label{BasicVortex} {\cal L}=-\frac{1}{4e^2}F_{\mu\nu}F^{\mu\nu}-
|(\de_{\mu}-iA_{\mu})q|^2-V(|q|)\ .
\end{equation}
The potential $V(|q|)$ is such that the field $q$ acquires a vev
$q_0$ and gives  mass to the $U(1)$ photon.  In this phase the
theory admits vortex solutions called Abrikosov-Nielsen-Olesen (ANO)
flux tubes \cite{Abrikosov,NielsenOlesen}. The ANO vortex is a
soliton extended in $1+1$ dimensions. The usual way is to choose
cylindrical coordinates $(z,r,\theta)$ with the vortex oriented in
the $\hat{z}$ direction. The fields can then be put in the following
form \bea \label{vortex} q &=& e^{i n \theta} q(r) \ ,
\\  A_{\theta} &=& \frac{n}{ r} A(r) \ , \nonumber \eea
where $q(r)$ and $A(r)$ are profile functions that must be
determined using the equation of motion. The field $q$ at $r \to
\infty$  has been chosen to lie in the vacuum manifolds $|q|=q_0$.
This is necessary to have a finite energy configuration. The
solution (\ref{vortex}) is obtained by choosing the element $n$ of
the homotopy group $[n]\in \pi_1(\mathbf{S}^1)$.  To have finite
energy it is also necessary to turn on a gauge field in order for the
covariant derivative $Dq$ to vanish. This creates a magnetic flux
that is exactly proportional to the topological number $n$.

\subsection{The Wall Vortex Limit} \label{wallim}

Now we describe what is the wall vortex and then we claim that
multi-vortices are wall vortices in the large $n$ limit.

\begin{figure}
 \begin{center}
\includegraphics[width=7 truecm]{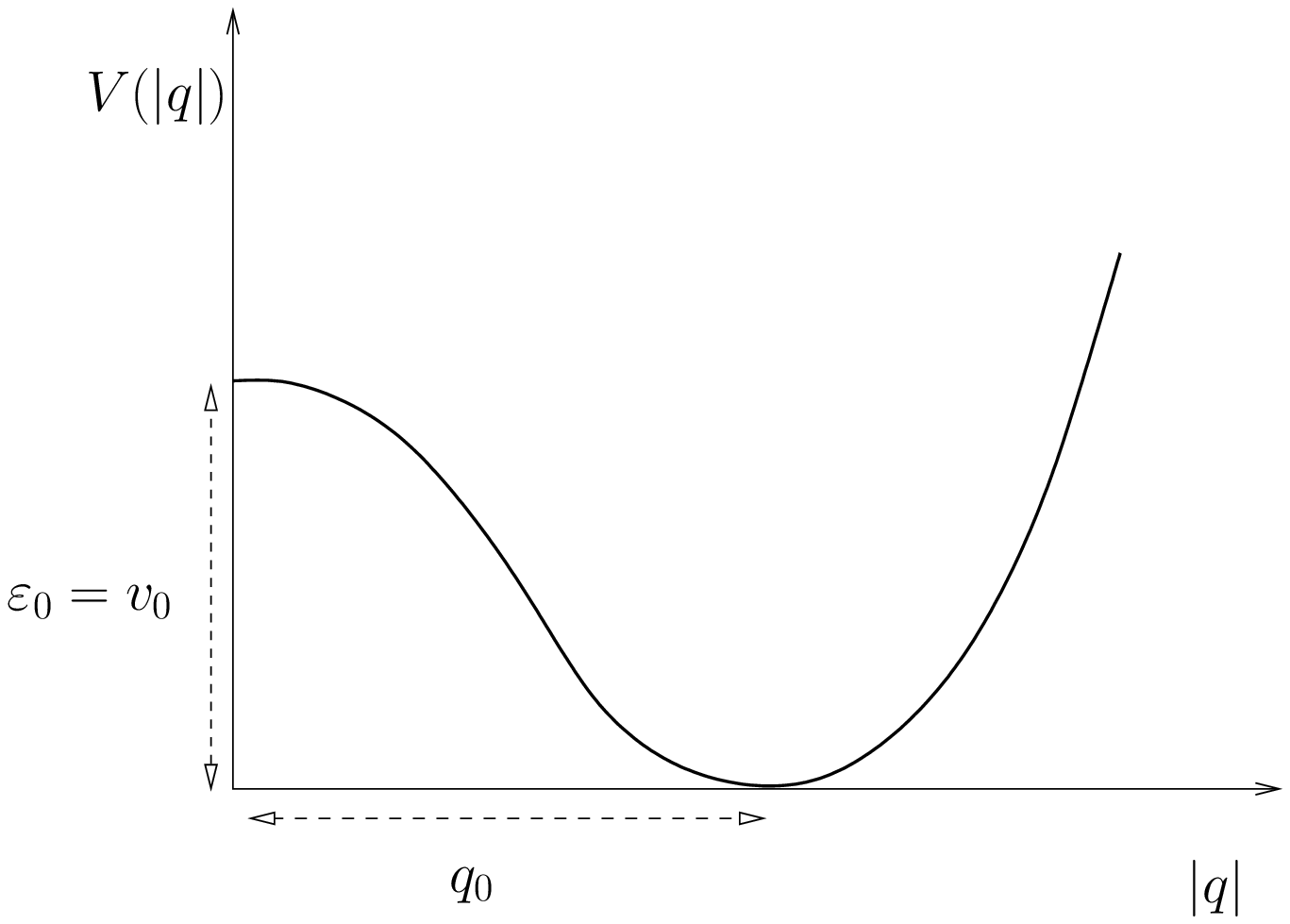}
\includegraphics[width=7 truecm]{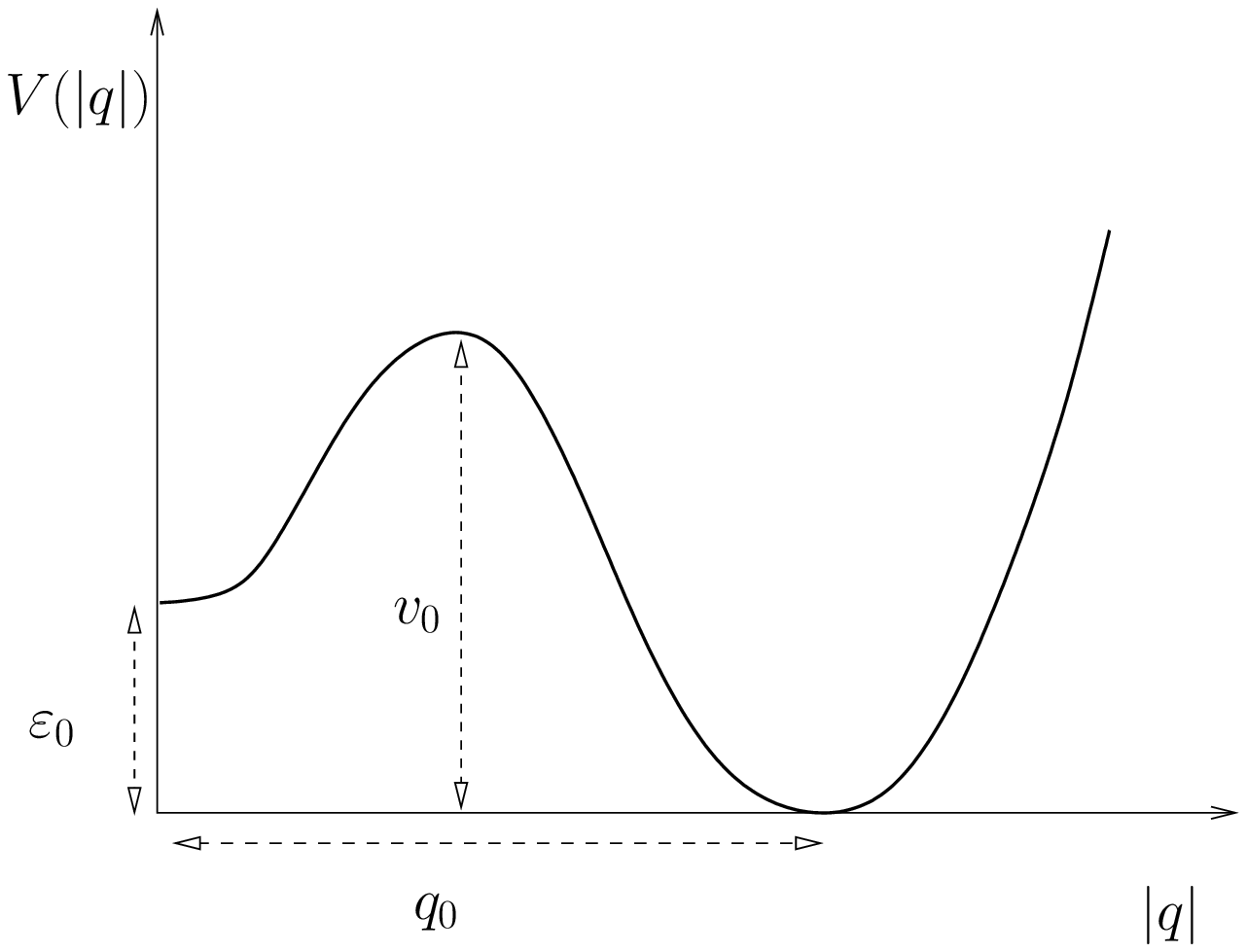}
\end{center}
\caption{\footnotesize Two possibilities for the Higgs
potential. In the first case the Coulomb phase is instable while in
the second case it is metastable.} \label{potenziali}
\end{figure}

The main idea is that the Coulomb phase $q=0$, even if it is not the
true vacuum of the theory, is always a stationary point of the
potential. This in fact is a trivial observation but, as we will
see, it is full of non-trivial consequences. In general there are
two cases described in Figure \ref{potenziali}.
The Coulomb phase is  a false vacuum with energy density
$V(0)=\eps_0$ and can be metastable or instable. We call $v_0$ the
highest peak of the potential between the Coulomb and the Higgs
phase.

Now comes the first assumption. Consider the domain wall
interpolating between the true and the false vacuum as an
independent object with tension and thickness, respectively: \beq
T_W \sim \sqrt{v_0}q_0\ , \qquad \Delta_W \sim
\frac{q_0}{\sqrt{v_0}} \ .\eeq This domain wall does not really
exist as an independent object. The energy density $\eps_0$ gives a
negative pressure that tends to eliminate the Coulomb half space
leaving only the Higgs vacuum. But, as we will see, it can exist in
the vortex solution where the negative pressure is counter balanced
by the pressure of the magnetic field. The key point is that the
domain wall maintain its ``identity'' even if we change the quanta
of magnetic flux $n$. Thus in the large $n$ limit the radius of the
vortex $R_V$ becomes large while the thickness of the wall
$\Delta_W$ remains fixed. This is the key assumption that needs to
be proven.

For the moment let's assume that the wall has its own identity and
then go on to the consequences.  The wall vortex is a flux tube
obtained in the following way. We compactify the domain wall on a
cylinder of radius $R$, keeping the Coulomb phase inside and the
Higgs phase outside. Then we turn on a magnetic flux inside the
cylinder and we write the tension of the tube as function of the
radius : \beq \label{unavariabile} T(R) = \frac{2 \pi n^2}{e^2 R^2}
+ T_W 2 \pi R + \varepsilon_0 \pi R^2 \ .\eeq There are three forces
that enter the game. Two of them tend to squeeze the tube and are
due to the tension of the wall and to the energy density $\eps_0$.
This is nothing but the Derrick collapse coming from the scalar part
of the Lagrangian. The last force is a positive pressure that comes
from the magnetic field inside the tube.

There are two regimes in which (\ref{unavariabile}) easily can be
solved: the SLAC bag and the MIT bag.\footnote{We have taken the
names from the bags models of hadrons: the MIT bag
\cite{Chodos:1974je} and the SLAC bag \cite{Bardeen:1974wr}. In the
Friedberg-Lee model \cite{Friedberg:1977xf} they arise as different
limits.}
\begin{description}
\item[SLAC bag:]
  This  regime is when $n$ satisfies the two conditions \beq
\label{surface} \frac{{q_0}^2 e}{ \sqrt{v_0}} \ll n \ll
\frac{{q_0}^2 e v_0}{{\varepsilon_0}^{3/2}} \ . \eeq The first
condition is that the radius $R_V$ is much bigger than the thickness
of the wall $\Delta_W$. The second condition is that the surface
term $T_W 2 \pi R$ dominates over the volume term $\varepsilon_0 \pi
R^2$. In this limit \beq \label{slacregime} T_{\textrm{SLAC}}=3
\sqrt[3]{2} \pi \, \left(\frac{ {T_W}}{e}\right)^{2/3} \, n^{2/3} \
, \qquad R_{\textrm{SLAC}} = \sqrt[3]{2} \, \frac{1}{ e^{2/3}
{T_W}^{1/3}} \, n^{2/3} \ . \eeq
  \item[MIT bag:] This  regime is when $n$ satisfies the condition
 \beq \frac{{q_0}^2 e v_0}{{\varepsilon_0}^{3/2}} \ll n\ . \eeq  In this limit the volume term
in (\ref{unavariabile}) dominates over the surface term and \beq
\label{volume} T_{\textrm{MIT}} = 2\sqrt{2 }\pi \, \frac{
\sqrt{\varepsilon_0}}{e} \, n  \ , \qquad R_{\textrm{MIT}} =
\sqrt[4]{2} \, \frac{1}{ e^{1/2} {\varepsilon_0}^{1/4}} \, \sqrt{n}
\ .\eeq Note that the tension is proportional to $n$, as in the BPS
case.
\end{description}

Let's write for clarity the complete story of multi-vortices. At the
value $n \gg \frac{{q_0}^2 e}{ \sqrt{v_0}}$, where the radius is
much bigger than the thickness of the wall, the multi-vortex becomes
a wall vortex. If the parameter $v_0$ is much bigger than $\eps_0$,
the wall vortex can be subdivided into two different regimes. In the
first one $ n \ll \frac{{q_0}^2 e v_0}{{\varepsilon_0}^{3/2}} $ and
the surface term dominates over the volume term and we call it the
SLAC bag regime. In this regime the tension scales like $n^{2/3}$.
When $ n \sim \frac{{q_0}^2 e v_0}{{\varepsilon_0}^{3/2}} $, the
volume term starts to be comparable to the surface term and thus a
second order phase transition between the SLAC bag and the MIT bag
takes place.  Note that the MIT bag regime, whenever $\varepsilon_0
\neq 0$, is always present and always dominates in the large $n$
limit.

Now lets write the conjecture for clarity:

{\it Consider the Abelian Higgs model (\ref{BasicVortex}) with a
general potential that has a true vacuum at $|q|=q_0\neq 0$ and a
Coulomb phase with energy density $V(0)=\varepsilon_0 \neq 0$. Call
$T_V(n)$ the tension of the vortex with $n$ units of magnetic flux.
The conjecture is that} \beq \label{firstformulation} \lim_{n \to
\infty} T_V(n) = T_{\textrm{MIT}}(n)
\ .\eeq


\subsection{The Differential Equations} \label{rescaling}

The differential equations for the profile functions of the vortex
(\ref{vortex}) are \bea \label{eqdiff} && \frac{\d^2 q}{\d
r^2}+\frac{1}{r}\frac{\d q}{\d r}-n^2 \frac{(1-A)^2}{r^2}q-\frac12
\frac{\delta V}{\delta q}=0 \ , \\ && \frac{\d^2 A}{\d
r^2}-\frac{1}{r}\frac{\d A}{\d r}+2 e^2(1-A)q^2=0 \ , \nonumber \eea
where $n$ is the winding number. We are looking for some limit of
the parameters so that the vortex really looks like a wrapped wall.
In this limit the profile functions should be:
\bea  \label{theta} q(r) &=& q_0 \, \theta_H(r-R_V)\ ,  \\
A(r)&=&r^2/R_V^2  \qquad  0 \leq r \leq R_V \ , \nonumber \\
A(r)&=&1\  \qquad  r \geq R_V \ , \nonumber \eea  where $\theta_H$
is the Heaviside step function. Note that the magnetic field is
$\propto A^{\prime}/r$ and is constant inside the vortex and zero
outside.

First of all we manipulate the differential equations (\ref{eqdiff})
to simplify them. The potential can be written as a dimensionless
function \beq \label{rescpot} V(q)=\varepsilon_0 \, \V
\left(\frac{q}{q_0}\right)\ , \eeq where $\varepsilon_0$ is the
value of the potential at $q=0$ and $q_0$ is the vev at which the
potential vanishes. The rescaled potential fullfills $\V(0)=1$ and
$\V(1)=0$.   We also rescale the scalar field $q=q_0 \, \chi$.
 In the case of a quartic potential the only one with the required properties is $\V(\chi)=(\chi^2-1)^2$.
 After these rescalings the equations
(\ref{eqdiff}) for the profiles become: \bea \label{eqscal1} &&
\frac{\d^2 \chi}{\d r^2}+\frac{1}{r}\frac{\d \chi}{\d r}-n^2
\frac{(1-A)^2}{r^2} \chi- a \frac{\delta \V}{\delta \chi}=0 \ , \\
\label{eqscal2} && \frac{\d^2 A}{\d r^2}-\frac{1}{r}\frac{\d A}{\d
r}+ b (1-A)\chi^2=0 \ . \eea There are three parameters that enter
the game: \beq n\ , \qquad a=\frac12 \frac{\varepsilon_0}{{q_0}^2}\
, \qquad b=2 e^2 {q_0}^2\ .\eeq

Now comes the first non-trivial hint for the wall vortex conjecture.
If the wall limit exists, then formula (\ref{volume}) can be trusted
in this limit. But the radius of the vortex $R_V$ comes from
equations (\ref{eqdiff}) and must depend only on the three relevant
parameters $n,a,b$. In general a function of $n,e,\varepsilon_0,q_0$
cannot be expressed as a function of $n,a,b$, but for (\ref{volume})
this is possible: \beq R_{\textrm{MIT}} = \sqrt[4]{2} \,
\frac{\sqrt{n}}{\sqrt[4]{a b}} \ . \eeq If we would not have found
such an expression, we would have concluded that the wall vortex
limit does not exist. This result encourages us to go on.

In this paragraph we look for a limit in which the radius of the
vortex remains constant and the solution approaches the wall vortex
(\ref{theta}). To do so we need to rescale also the parameters $a$
and $b$ such that the radius remains fixed while approaching the
large $n$ limit. The $\chi$ profile must become a step function: it
is zero inside $R_V$, it goes from zero to one in a distance
$\Delta_W$, and then remains constant at one. Thus
$\chi^{\prime\prime}$ in equation (\ref{eqscal1}) develops a
$\delta'(r-R_V)$ singularity or, in terms of $\Delta_W$, $\chi''
\sim 1/{\Delta_W}^2$. To counter balance this divergence in
(\ref{eqscal1}) we must have that $a$ goes to infinity like
$1/{{\Delta_W}^2}$. Now consider the second equation (\ref{eqscal2})
where $A''(R_V)$ has a $(2 / R_V) \delta(r-R_V) $ singularity, or in
terms of $\Delta_W$, $A''(R_V) \sim 1/(R_V \Delta_W)$. Since
$(1-A)\chi$ is of order $\Delta_W / R_V$ around $R_V$, we must also
send $b$ to infinity like $1/{\Delta_W}^2$.
 Now we can reformulate the
conjecture of the wall limit in another equivalent way.

\textit{ Consider the succession of parameters $a_n=n a_1$  and $
b_n=n b_1$ and name the solution of (\ref{eqscal1}) and
(\ref{eqscal2}) with the vortex boundary conditions,
$\chi_{n,a_n,b_n}(r)$ and $A_{n,a_n,b_n}(r)$. In the limit $n \to
\infty$} \bea \label{reformulation}  &&
\lim_{n \to \infty} \chi_{n,a_n,b_n}(r) \to \theta_H(r-R_V)\ , \\
&& \lim_{n \to \infty} A_{n,a_n,b_n}(r) \to \left\{
\begin{array}{cc} r^2/{R_V}^2 & 0 \leq r \leq R_V \ , \\
1 & r>R_V  \ . \\ \end{array}\right. \nonumber \eea

 This limit
has been chosen such that the radius of the vortex remains constant
$R_V = \sqrt[4]{2/(a_1 b_1)}$ and also the ratio $a_n / b_n$ remains
constant. The information about the ratio $a / b$ disappears in the
wall vortex limit. It is only related to the shape of the limiting
functions $\chi_{n,a_n,b_n}(r)$ and $A_{n,a_n,b_n}(r)$. Probably a
stronger version of the conjecture is  true:  the ratio $a_n /b_n$
is kept limited from above and from below during the limit, so that
it does not go neither to infinity nor to zero.

The ratio $a/b$ has also another important meaning. In fact, in the
case of quartic potential, it is essentially the parameter $\beta$
that measures the ratio between the Higgs and the photon mass: \beq
\beta=\frac{m_{H}}{m_{\gamma}}=8\frac{a}{b} \ .\eeq When $\beta <1 $
the Higgs attraction dominates and the vortices are of Type I. When
$\beta>1$ the photon repulsion dominates and the vortices are of Type
II. When $\beta=1$ the vortices are BPS, this means that the tension
is a linear function of $n$ and the distance between different vortices
is the vortex moduli space.

\subsection{A Non Trivial Check}

Now we make a non-trivial check of the result (\ref{volume}) using
the famous example solved by Bogomol'nyi \cite{Bogomolny}. When the
potential is \beq \label{BPSpotential}
V(|q|)=\frac{e^2}{2}(|\phi|^2-\xi)^2 \ ,\eeq the tension is \beq
\label{BPStension} T_{\textrm{BPS}}=2 \pi n \xi \ , \eeq for all
$n$. Solving the model with our trick, the result must coincide with
eq. (\ref{BPStension}).  For the BPS potential (\ref{BPSpotential}),
the energy density of the instable Coulomb vacuum is
$\varepsilon_0=e^2\xi^2/2$. Using (\ref{volume}),  we find exactly
(\ref{BPStension}). This could hardly be just a coincidence.

It is well known that $n$ BPS vortices have a moduli space of real
dimension $2n$ where the coordinates can be interpreted as the
position of every constituent $1$-vortex \cite{modilispace}. In the
BPS bag we expect to recover this moduli space and we expect it to
have infinite dimension. In fact this space consist of the closed
surfaces with fixed area. More details will be discussed in
\cite{monopolemio} in relation with the multi-monopole moduli space.

\section{Numerical Analysis} \label{numerical}

This is the central part of the paper. With the help of numerical
computations we are going to prove the conjecture made in the
previous section. The strategy is to prove the conjecture as it has
been reformulated in (\ref{reformulation}), that is, we rescale the
parameters of the theory such that the radius of the vortex remains
constant and the thickness of the wall goes to zero.

The code\footnote{The code is written in the mathematical language
Maple.} used to solve the system of differential equations
(\ref{eqscal1}-\ref{eqscal2})  uses a numerical method called finite
difference method. The  boundary conditions are at the singular
points of the system $0$ and $\infty$ and thus we use the limits:
\bea \lim_{r\to
0}\,(2A(r)-rA^{\prime})=0 \ , &\qquad& \lim_{r\to 0} \, (n\chi(r)-r\chi^{\prime})=0 \ , \nonumber \\
\lim_{r\to\infty} \,
\left(A(r)+\frac{1}{\sqrt{nb}}A^{\prime}\right)=1 \ , &\qquad&
\lim_{r\to\infty} \,
\left(\chi(r)+\frac{1}{\sqrt{8na}}\chi^{\prime}\right)=1
\label{numlastbc} \ , \eea  Other methods often used are e.g. the
shooting method, but in case of a step function insanely high
accuracy is needed. Still for the finite difference method, the step
function poses severe problems and the technique to deal with the
problem is to use a combination of a continuation method and feeding
the algorithm with an approximate solution. Also the rescaling of
the equations such that the radius of the vortex is constant is an
important factor for solving the equations.
 The finite
difference method transforms the problem into a matrix equation \beq
F_{ij} = 0 \ , \quad i=1,\ldots,\textrm{meshsize} \ ,\quad j=1,2 \ ,
\eeq which, when using a continuation method, looks like \beq
G_{ij}(t) = 0 \ ,\quad t\in[0,1]\ ,\eeq where the problem is easy to
solve when $t=0$ and hard to solve for $t=1$. Then the solver will
try to increase the hardness in sufficiently small steps until the
final problem $t=1$ is solved. Whenever a solution has been found it
is fed to the solver as an approximate solution which is the
solution to the problem for $t=0$ and for sufficiently small steps
it is then possible to achieve higher winding number $n$. The
approximate solutions tell the solver where to look for the solution
in phase space.

\subsection{Quartic Potential and the MIT bag regime}

The simplest case is the quartic potential. After the rescaling
(\ref{rescpot}) there is only one quartic potential to be
considered:  \beq \V(\chi)=(\chi^2-1)^2 \eeq

\begin{figure}[!p]
\begin{center}
\includegraphics[width=7.5 truecm]{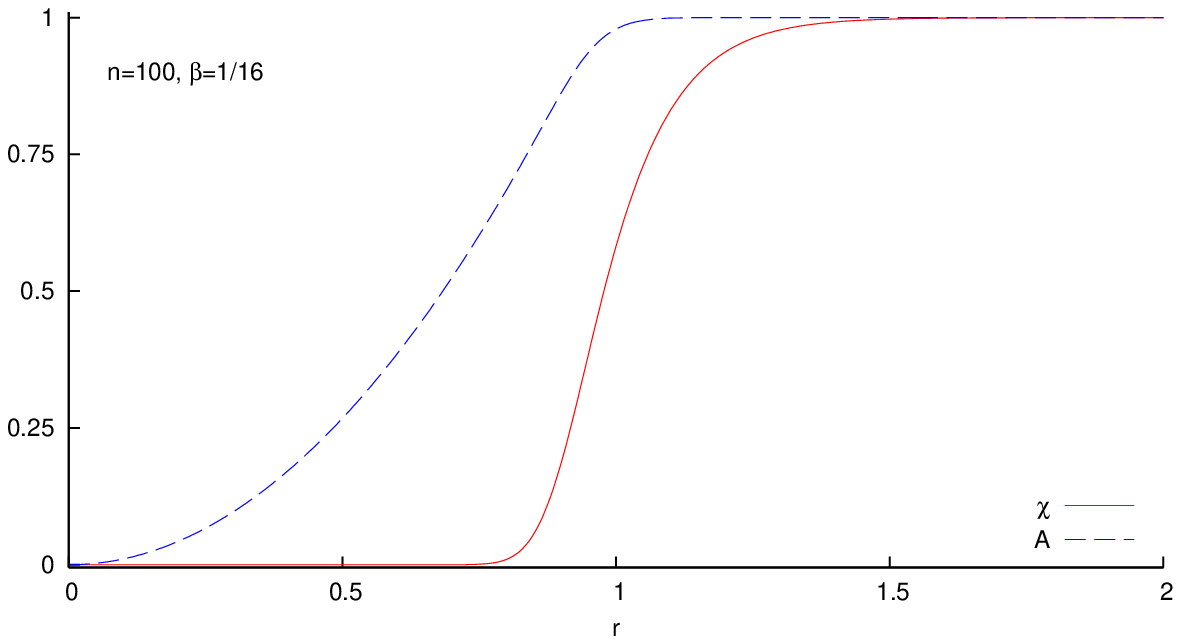}
\includegraphics[width=7.5 truecm]{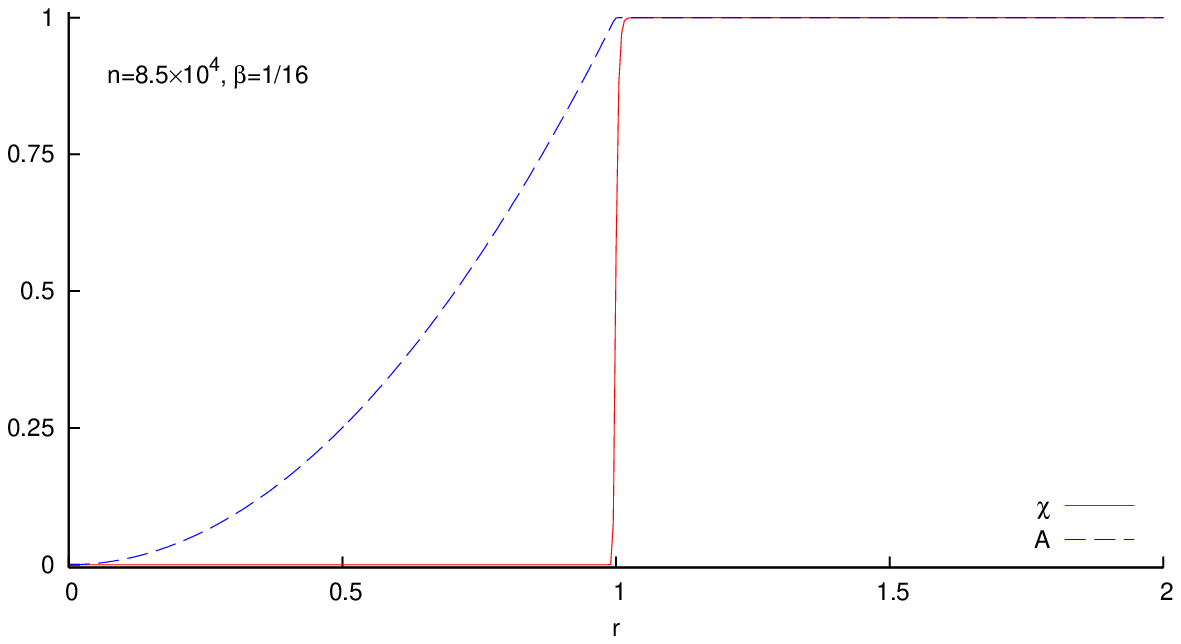}
\end{center}
\caption{\footnotesize$\chi(r)$ and $A(r)$ profiles for $\beta=1/16$
and $n=100$, $n=85,000$. } \label{profilibetaless}
\end{figure}

\begin{figure}[!p]
\begin{center}
\includegraphics[width=7.5 truecm]{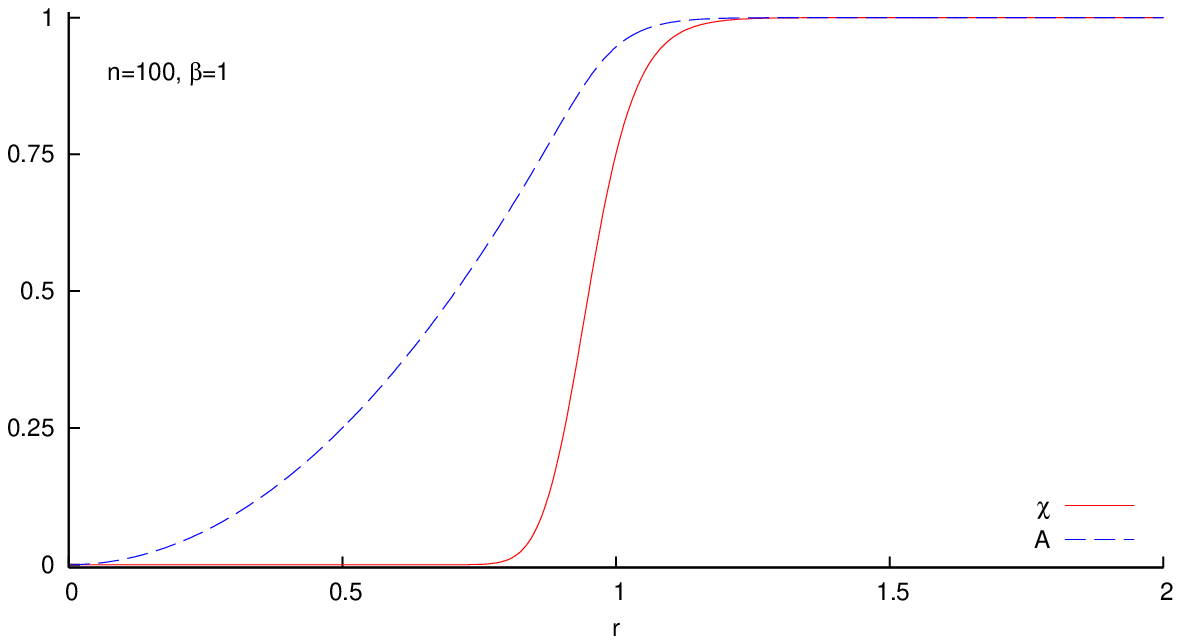}
\includegraphics[width=7.5 truecm]{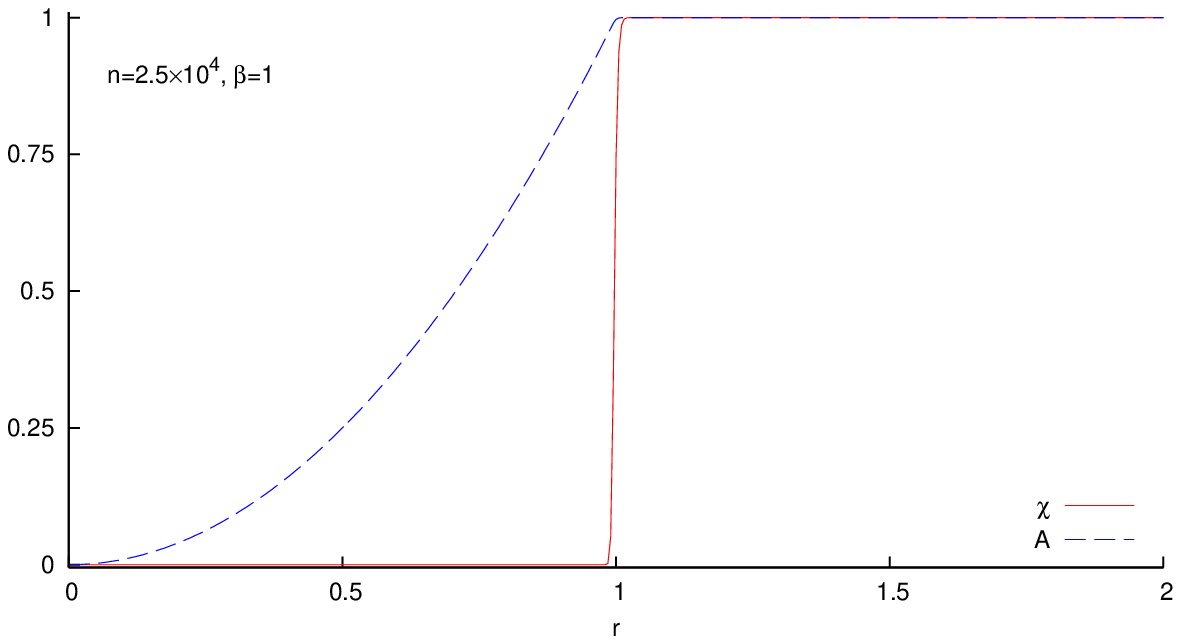}
\end{center}
\caption{\footnotesize$\chi(r)$ and $A(r)$ profiles for $\beta=1$
and $n=100$, $n=25,000$.  } \label{profilibetaBPS}
\end{figure}

\begin{figure}[!p]
\begin{center}
\includegraphics[width=7.5 truecm]{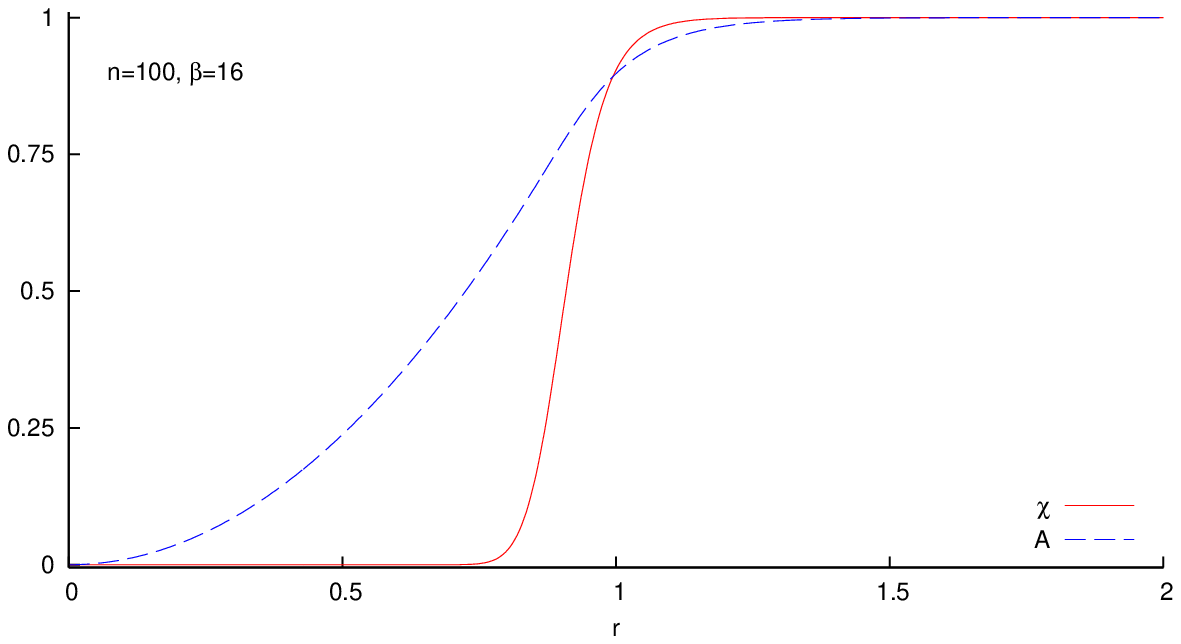}
\includegraphics[width=7.5 truecm]{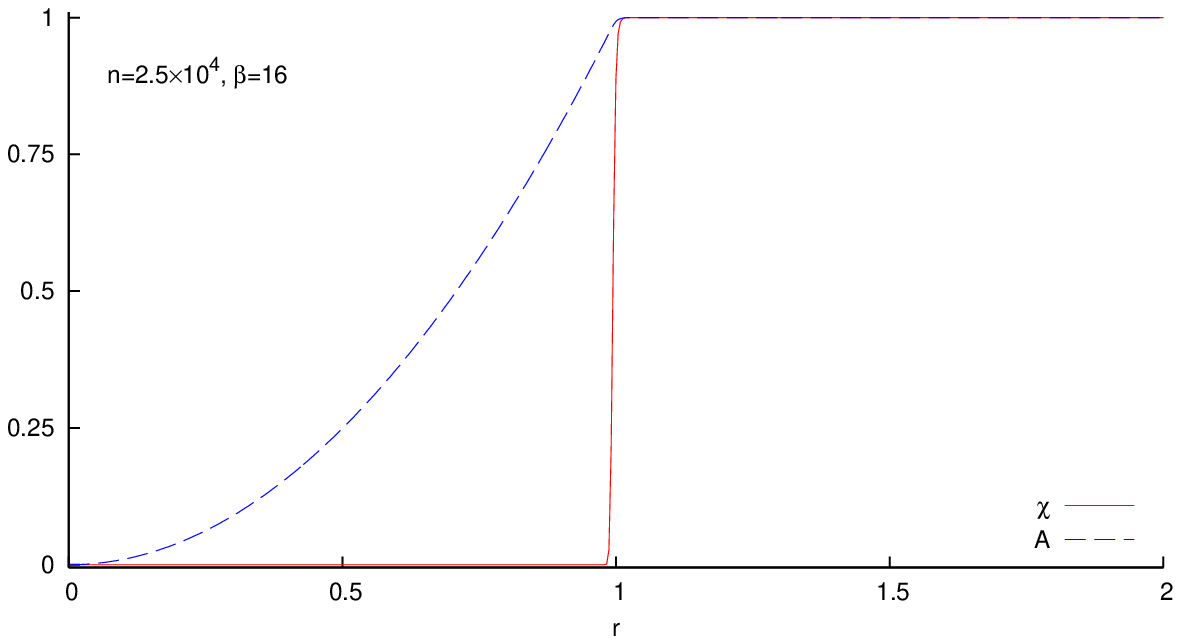}
\end{center}
\caption{\footnotesize$\chi(r)$ and $A(r)$ profiles for $\beta=16$
and $n=100$, $n=25,000$. } \label{profilibetamore}
\end{figure}

The following figures show the results of the numerical program.
Figure \ref{profilibetaBPS} is the BPS case where $\beta=1$, Figure
\ref{profilibetaless} is the type I case with $\beta=1/16$ and
Figure \ref{profilibetamore} is the type II case with $\beta=16$. In
every case we present the profiles for small $n=100$ and large $n$.
At $n=25,000$ the step function for $\chi(r)$ is evident. Note that
in the large $n$ limit the information of $\beta$ is almost lost and
the profiles are all the same. The case $n=100$ is interesting
because we can see the difference among BPS, type I and type II
while approaching the wall vortex limit.

As another check we now evaluate the tension of the vortex. If we
directly insert the ansatz (\ref{vortex}) into the energy density
derived from the Lagrangian (\ref{BasicVortex}), we obtain: \beq
T_V(n) =2\pi \int r dr \Big[\frac{1}{2e^2}\Big(\frac{n
A^{\prime}}{r}\Big)^2  +\frac{n^2}{r^2}(1-A)^2 q^2
+{q^{\prime}}^2+V(q)\Big] \ .\eeq Now we want to verify the
conjecture in its first formulation (\ref{firstformulation}), that
is by increasing $n$ without changing the parameters of the theory.
We introduce a new quantity ${\cal T}(n)$ defined to be the ratio
between the real tension and the BPS tension: \beq {\cal
T}(n)=\frac{T_V(n)}{T_{\textrm{BPS}}(n)}\eeq For this new quantity
the conjecture (\ref{firstformulation}) is simply $\lim_{n \to
\infty}{\cal T}(n) =1 $.

\begin{figure}[h!t]
\begin{center}
\includegraphics[width=9 truecm]{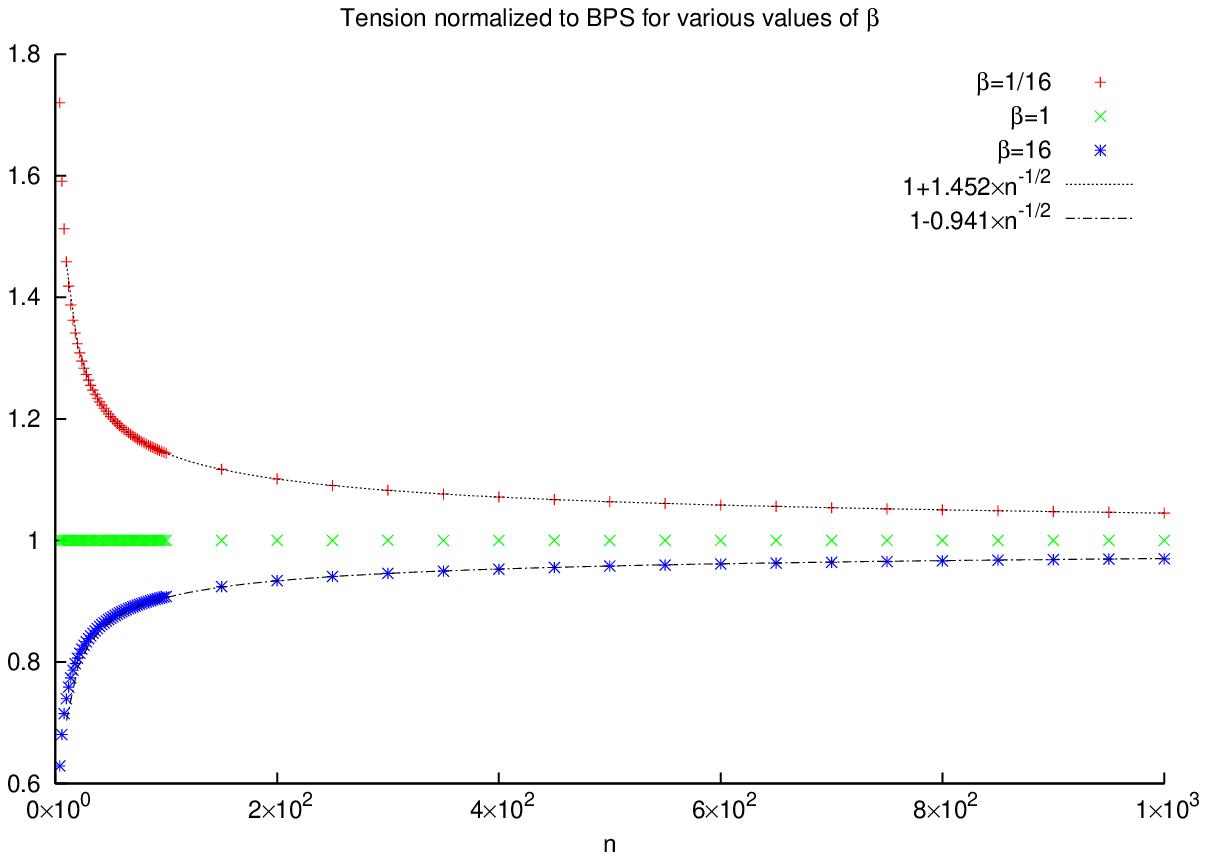}
\end{center}
\caption{\footnotesize Plot of the normalized tension ${\cal T}$ as
function of $n$ for the three cases $\beta=1/16,1,16$.}
\label{tensionlinreg}
\end{figure}

Figure \ref{tensionlinreg} shows the three plots of ${\cal T}(n)$
for $\beta=1/16,1,16$, respectively. The quantity ${\cal T}(n)$ has
also an important physical meaning: it is the tension per unit of
flux carried by the vortex. This means that from the sign of the
derivative $\d {\cal T}/\d n$, we can read if there is attraction or
repulsion between vortices. Figure \ref{tensionlinreg} is consistent
with the ordinary expectation of type I and type II superconducting
vortices. If we take for example $\beta =1/16$ (type I vortices) the
derivative $\d {\cal T}/\d n$ is negative and this means that there
is attraction. For $\beta = 16$ the function ${\cal T}$ grows up to
$1$ and this means that there is repulsion (type II vortices). Our
theory predicts also the way ${\cal T}(n)$ approaches $1$ at
infinity. The radius of the vortex in the large $n$ limit is
$R_{\textrm{MIT}} \sim \sqrt{n}$ while the thickness of the wall
$\Delta_W$ remains constant. This mean that the deviation from the
``perfect'' wall vortex is of order $1/\sqrt{n}$. In Figure
\ref{tensionlinreg} we show also the fits ${\cal T}(n) \sim 1 +
\textrm{const}/\sqrt{n}$ and they perfectly agree with the numerical
data.
\begin{figure}[h!t]
\begin{center}
\includegraphics[width=9 truecm]{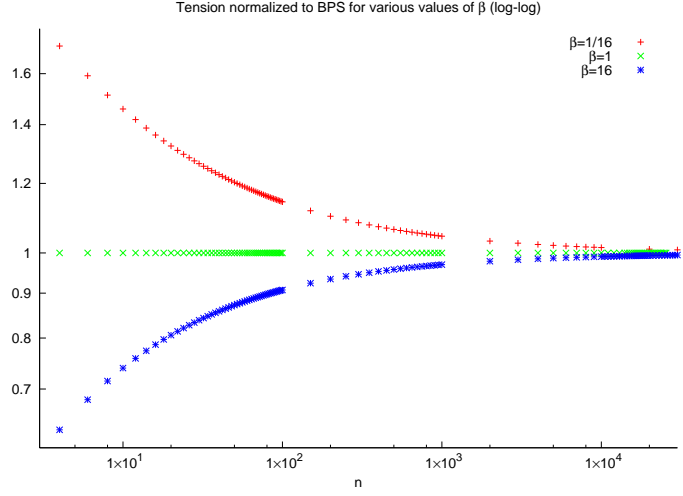}
\end{center}
\caption{\footnotesize Plot of the normalized tension ${\cal T}$ in
the log-log graph.} \label{tensionlog}
\end{figure}

\subsection{Degenerate Vacua and the SLAC bag regime\label{degenerate}}

\begin{figure}[h!t]
\begin{center}
\includegraphics[width=8 truecm]{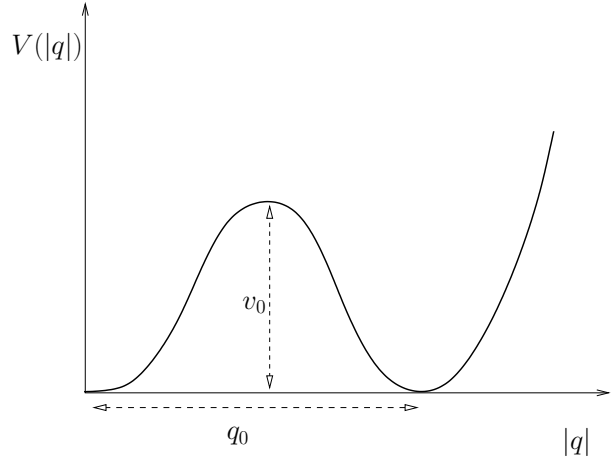}
\end{center}
\caption{\footnotesize The Coulomb phase and the Higgs phase
are the two degenerate vacua of the potential.} \label{degeneratepot}
\end{figure}

Now we consider the case of degenerate vacua when the potential is
like that of Figure \ref{degeneratepot}.
Since the Coulomb vacuum energy density vanishes, there is no MIT
regime in the large $n$ limit. The conjecture
(\ref{firstformulation}) must then be replaced by \beq
\label{spectrum} \lim_{n \to \infty} T_V(n) = T_{\textrm{SLAC}}(n)
\ . \eeq

If we want to test the conjecture with our numerical code, it is
necessary to find a rescaling of the parameters so that the radius
of the vortex remains fixed while the thickness of the wall
converges to zero. Since the large $n$ limit is a SLAC bag regime,
we must properly modify all the analysis that we have done in
subsection \ref{rescaling}. This has already been done in
\cite{wallandfluxes} so we present only a brief description. The
equation (\ref{rescpot}) must now be substituted with \beq
\label{rescpot2} V(q)=v_0 \V \left(\frac{q}{q_0}\right)\ . \eeq The
differential equations (\ref{eqscal1}-\ref{eqscal2}) are the same
and the three relevant
 parameters are
\beq n\ , \qquad a=\frac12 \frac{v_0}{{q_0}^2}\ , \qquad b=2 e^2
{q_0}^2\ .\eeq

The radius of the SLAC bag (\ref{slacregime}) can be rewritten in
terms of the three relevant parameters \beq R_{\textrm{SLAC}} \sim
\frac{n^{2/3}} {a^{1/6} b^{1/3}} \ . \eeq We can now give the new
formulation of the conjecture.

{\it Consider the succession of parameters $a_n=n^{4/3} a_1$ and $
b_n=n^{4/3} b_1$ and call the solution of (\ref{eqscal1}) and
(\ref{eqscal2}) with the vortex boundary conditions,
$\chi_{n,a_n,b_n}(r)$ and $A_{n,a_n,b_n}(r)$. In the limit $n \to
\infty$} \bea && \lim_{n \to \infty} \chi_{n,a_n,b_n}(r) \to
\theta_H(r-R_v)\ , \\ && \lim_{n \to \infty} A_{n,a_n,b_n}(r) \to
\left\{
\begin{array}{cc} r^2/{R_V}^2 & 0 \leq r \leq R_V \ , \\
1 & r>R_V  \ . \\ \end{array}\right. \nonumber \eea   This limit has
been chosen so that the radius of the vortex remains constant $R_V
\sim n^{2/3} {a_n}^{-1/3} {b_n}^{-1/6}={a_1}^{-1/3} {b_1}^{-1/6}$
and also the ratio $a_n / b_n$ remains constant. The ratio $a / b$
disappears in the wall limit and is only related to the shape of the
limiting functions $\chi_{n,a_n,b_n}(r)$ and $A_{n,a_n,b_n}(r)$.

\begin{figure}[h!t]
\begin{center}
\includegraphics[width=9 truecm]{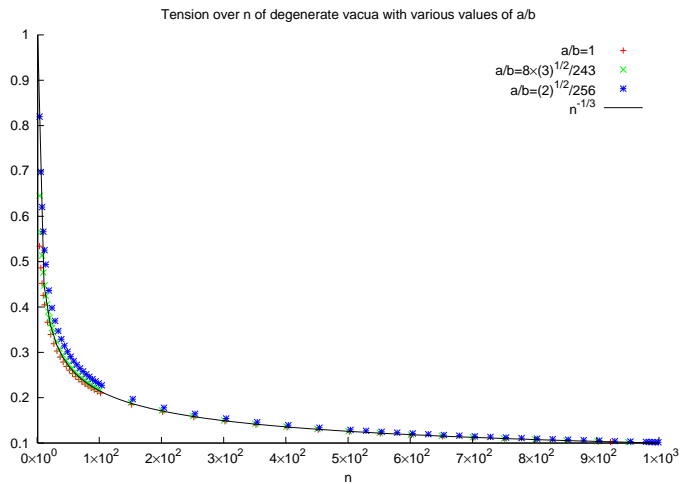}
\end{center}
\caption{\footnotesize Tension per unit of flux for three
different values of $a/b$. At large $n$, ${\cal T} \propto n^{-1/3}$
as predicted by the theory.} \label{degeneratetovern}
\end{figure}

For the computation we take the simplest potential with the
Coulomb and Higgs degenerate vacua:
 \beq \label{degpot}\V(\chi)=\chi^2(\chi^2-1)^2 \ .\eeq
In Figure \ref{degeneratetovern} we have plotted the tension
per unit of flux $T(n)/n$ for three different values of $a/b$. As
predicted by the theory all the lines converge to $T/n \propto
n^{-1/3}$ as $n$ becomes large.
\begin{figure}[h!t]
\begin{center}
\includegraphics[width=9 truecm]{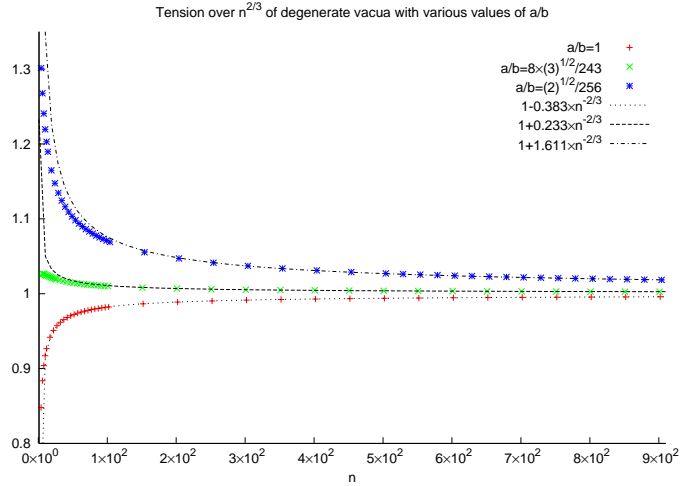}
\end{center}
\caption{\footnotesize Tension divided by $n^{2/3}$ for three
different values of $a/b$. At large $n$ they approach a constant
with an error of order $1/n^{2/3}$. }
\label{degeneratetoverntwothird}
\end{figure}

In Figure \ref{degeneratetoverntwothird} we have instead plotted
$T(n)/n^{2/3}$. In this way is more easy to see the SLAC regime
since at infinity all the three lines approaches a constant. In this
plot we can also see the first order deviation from the ``perfect''
SLAC bag. Since the radius $R_{\textrm{SLAC}} \propto n^{2/3}$ while
the thickness of the wall remains constant, we expect a deviation of
order $1/n^{2/3}$. The plots with corresponding fits confirm this
expectation.

An interesting question is if there exists a kind of BPS notion also
for the degenerate vacua potential. Maybe there do exist some
potential and some particular value of $a/b$ for which the tension
not only approaches $T_{\textrm{SLAC}}$ as $n$ goes to infinity but
is exactly equal to it for all values of $n$.

\subsection{SLAC/MIT phase transition}

Another thing to be checked is the presence of a window, namely the
SLAC regime, when the Coulomb vacuum is metastable and the peak
$v_0$ is much greater than the energy density $\varepsilon_0$. To
obtain this regime we need a potential like the second of Figure
\ref{potenziali} with $v_0 \gg \varepsilon_0$. The theoretical
analysis predicts that in the windows of parameters (\ref{surface})
the tension and the radius scale like eq.~(\ref{slacregime}).

The simplest way to obtain such a potential is to add an opportune
$\chi^6$ interaction:
 \beq \label{pot}\V(\chi)=(\sigma \chi^2+1)
(\chi^2-1)^2 \eeq The conditions $\V(0)=1$ and $\V(1)=0$ leave one
free parameter $\sigma$ that essentially measures the height of the
peak of the potential. When $\sigma$ is great enough the peak is at
$\chi=1/\sqrt{3}$ and has the value $v_0=4 \sigma /27$.

\begin{figure}[h!t]
\begin{center}
\includegraphics[width=6.5 truecm]{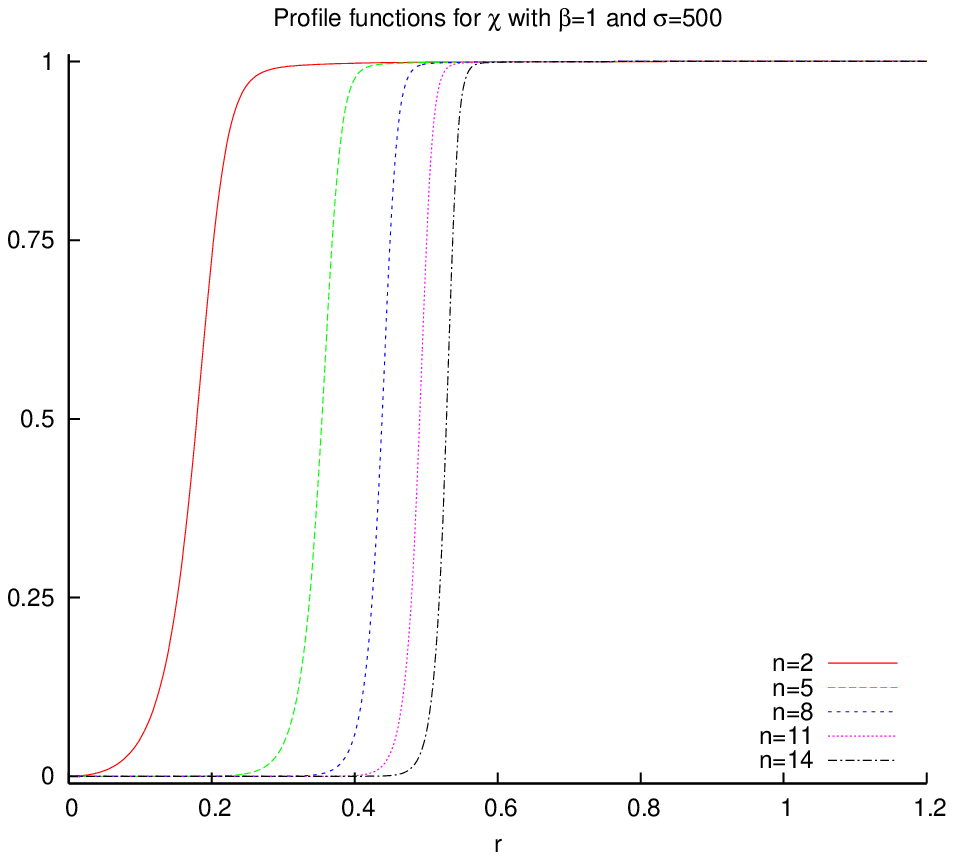}
\includegraphics[width=6.5 truecm]{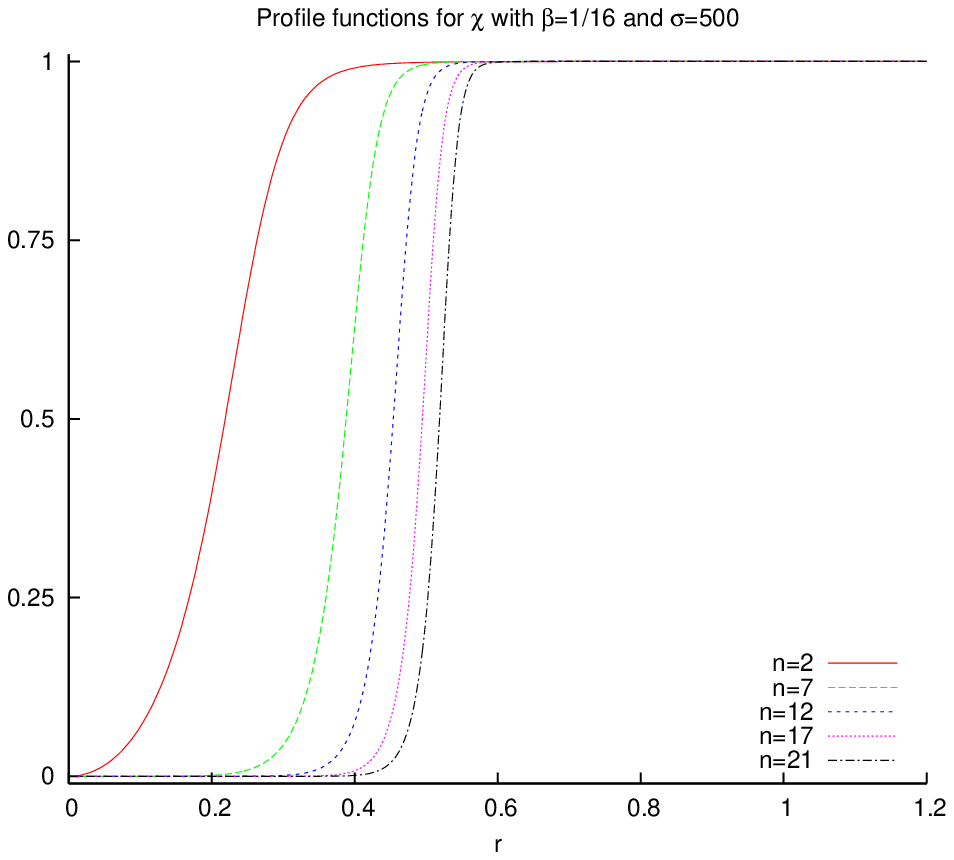}
\end{center}
\caption{\footnotesize Profile $\chi(r)$ for the potential
(\ref{pot}) with $\sigma=500$. The two graphs correspond
to $\beta=1$ and $\beta=1/16$, respectively.}
\label{sigmafivehundred}
\end{figure}

We need quite a big $\sigma$ to detect the window (\ref{surface}).
In Figure \ref{sigmafivehundred} we have plotted the profile
$\chi(r)$ for $\sigma = 500$. For this computation we have used the
rescaling $a_n=n a_1$, $b_n= n b_1$. In this rescaling the large $n$
limit should be a step function with constant radius
$R_{\textrm{MIT}}=1$. What we should see, to confirm the theory, is
that the $\chi(r)$ profile becomes a step function much before
reaching the MIT regime. The radius of the wall vortex should be
governed by the SLAC law and should increase as $n^{1/3}$. When the
radius reaches $1$ there should be a phase transition between the
SLAC regime and the MIT regime where the radius remains constant as
$1$.

Using the rescaling $a_n=n a_1$, $b_n= n b_1$ the numerical
computation cannot reach high enough $n$ to reach the phase
transition. Nevertheless it is possible to detect the presence of
the SLAC regime just by looking at Figure \ref{sigmafivehundred}.
The profile $\chi(r)$ becomes a step function at a radius much lower
than $1$ and then the radius increases towards $1$.

\begin{figure}[h!t]
\begin{center}
\includegraphics[width=9 truecm]{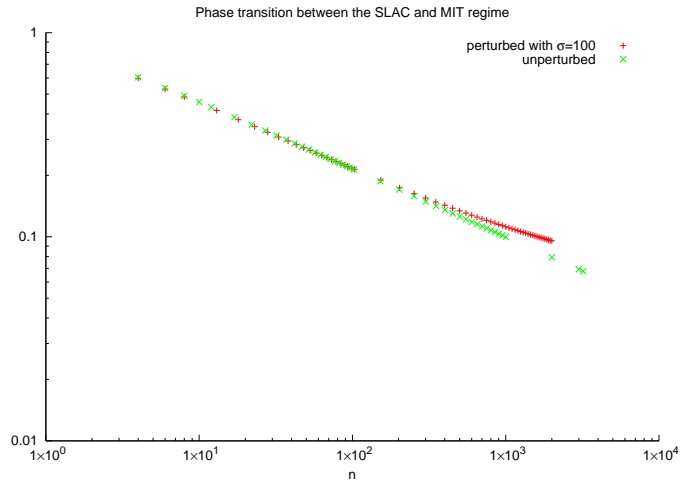}
\end{center}
\caption{\footnotesize In this graph it is possible to see the
phase transition between the SLAC regime and the MIT regime. The
green points correspond to the degenerate vacua potential that is
always in the SLAC regime. The red points correspond to the
perturbed potential (\ref{perturbed}) with $\sigma = 100$. At $n
\sim 10^2$ there is a second order phase transition.}
\label{phasetransition}
\end{figure}

To see the SLAC/MIT phase transition we have to use another
strategy. We use the rescaling adapted to the degenerate vacua
situation $a_n = n^{4/3} a_1$, $b_n = n^{4/3} b_1$ and we take the
potential to be \beq \label{perturbed}
\V(\chi)=(\chi^2+\frac{1}{\sigma}) (\chi^2-1)^2 \eeq so that a high
value of $\sigma$ corresponds to a small perturbation of the
degenerate vacua potential (\ref{degpot}). With this strategy it
will be possible to detect the phase transition. In Figure
\ref{phasetransition} we have plotted the tension for the degenerate
vacua potential (\ref{degpot}) and the perturbed degenerate vacua
potential (\ref{perturbed}) with $\sigma = 100$. The tensions are
the same up to $n \sim 10^2$ and then the perturbed potential starts
to deviate from the unperturbed one. Unfortunately the computation
stops here and we cannot see that it enters in the MIT bag phase.
Anyway the phase transition is clearly visible from the Figure.

\section{Conclusion and Further Developments} \label{discussion}

One of the most surprising aspects of the large $n$ limit
(\ref{firstformulation}), is that the tension depends only on the
value of the potential at zero $V(0)=\varepsilon_0$. If we have two
different potentials with the same $\varepsilon_0$, the large $n$
limit of the tension is the same.

Now we are going to argue something more. If we take two different
potentials $V_1(q)$ and $V_2(q)$ with the same $\varepsilon_0$ and
the same $q_0$, not only the tensions are the same in the large $n$
limit, but also are the profile functions.

The energy density derived from the Lagrangian (\ref{BasicVortex})
is the following \beq T_i[A,q] =2\pi \int r dr
\Big[\frac{1}{2e^2}\Big(\frac{n A^{\prime}}{r}\Big)^2
+\frac{n^2}{r^2}(1-A)^2 q^2 +{q^{\prime}}^2+V_i(q)\Big] \ ,\eeq
where the index $i=1,2$ refers to the potentials $V_1$ and $V_2$,
respectively. The energy density must be regarded as a functional:
it takes the two profile function $A(r)$ and $q(r)$ and it gives out
a number. The minimization of $T_i[A,q]$ gives the profiles and the
tension of the vortex. Now call $A_1(r)$ and $q_1(r)$ the profiles
obtained by the minimization procedure for the potential $V_1$, and
put them into the functional of the second potential. What we obtain
is \beq T_2[A_1,q_1]=T_1[A_1,q_1]+2\pi \int r dr
\Big[V_2(q_1)-V_1(q_1)\Big] \ , \eeq where we have simply added and
subtracted $V_1$. In the large $n$ limit the profile $q_1(r)$
becomes a step function and, since the two potentials have the same
$\varepsilon_0$ and $q_0$, the extra piece $2\pi \int r dr
[V_2(q_1)-V_1(q_1)]$ vanishes. This means that the functions
$A_1(r)$ and $q_1(r)$, being the minima of the functional $T_1$,
become also approximately the minima of the functional $T_2$.

 {\it If the potential $V_1(q)$
and $V_2(q)$ have the same $\varepsilon_0$ and the same $q_0$, the
profile functions converge in the large $n$ limit} \beq \lim_{n\to
\infty} ||q_1-q_2|| = 0 \ , \qquad \lim_{n\to\infty} ||A_1-A_2||=0 \
,\eeq {\it where the distance between functions is measured with the
$L^2$ metric.}
\begin{figure}[h!t]
\begin{center}
\includegraphics[width=7 truecm]{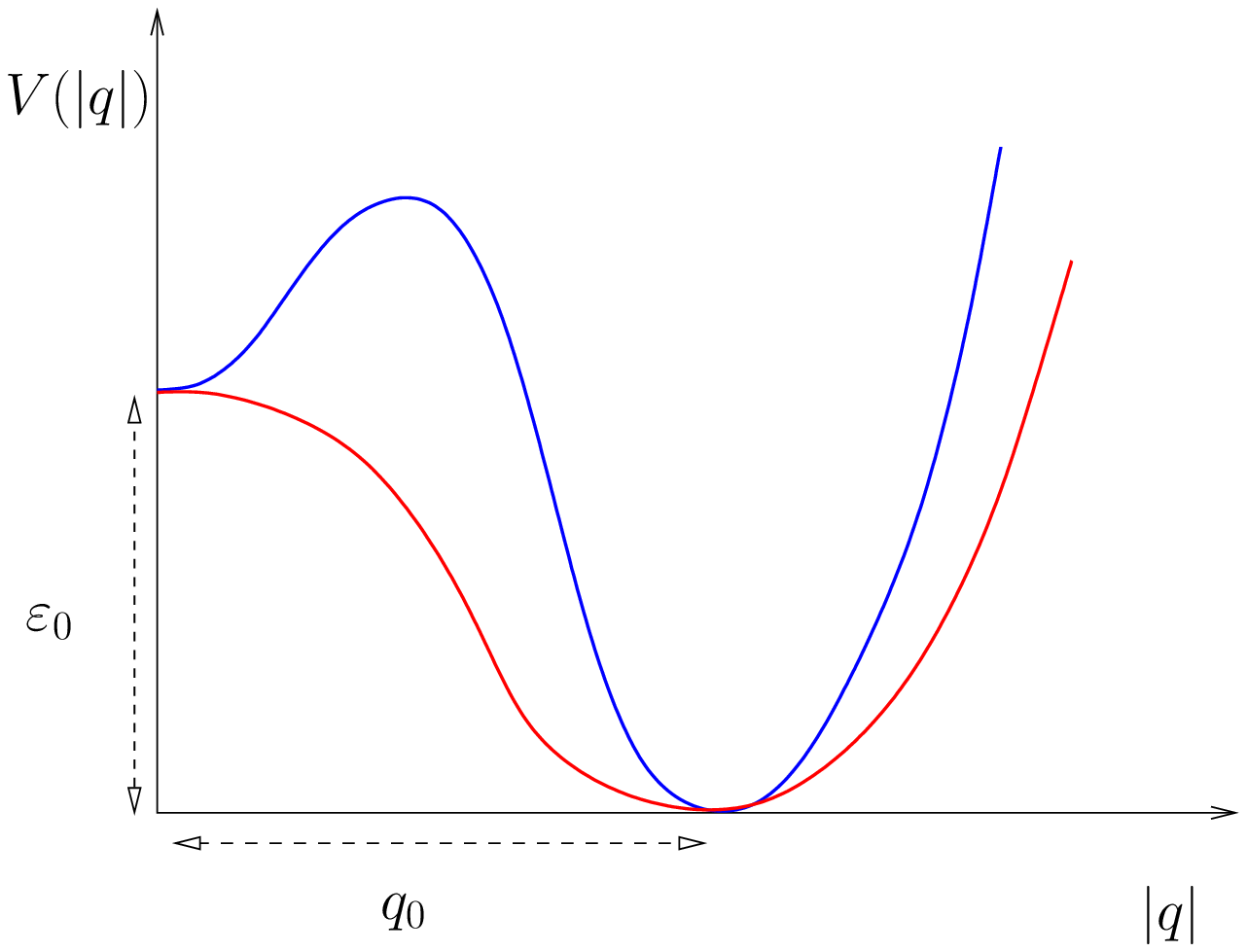}
\includegraphics[width=7 truecm]{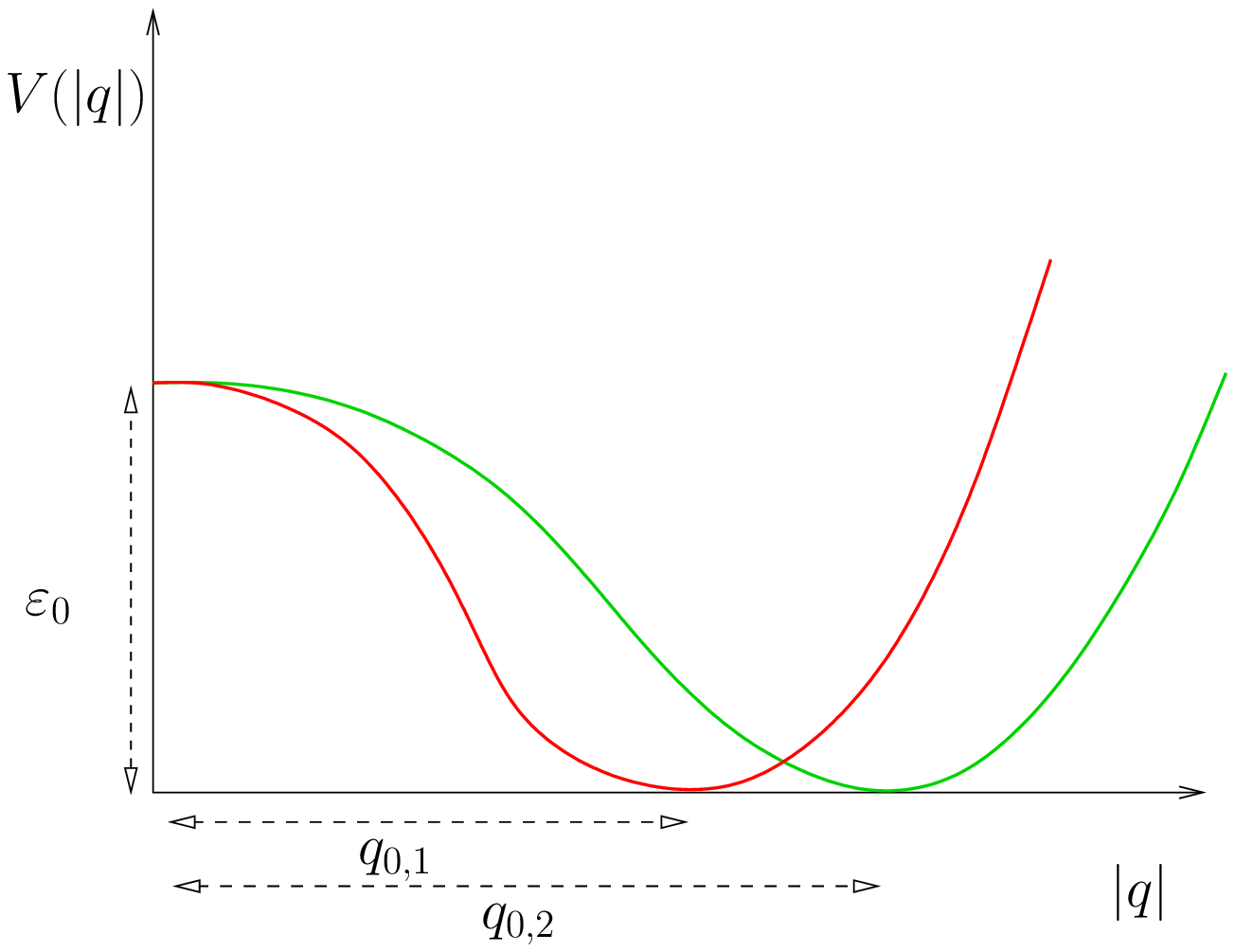}
\end{center}
\caption{\footnotesize In the first sketch we have two
different potentials with the same $\varepsilon_0$ and the same
$q_0$. In the second sketch we have two potentials with the same
$\varepsilon_0$ but different $q_0$.} \label{potentialcompno}
\end{figure}
If the conjecture is true we have also the following result.  If the
ratio between the zero energy density $\varepsilon_0$ and the vev
$q_0$ is the same of the BPS potential, that is $\varepsilon_0=e q_0
/2$, whatever the shape of the potential, we recover supersymmetry
in the large $n$ limit. It is in fact known that the Abelian-Higgs
model with the BPS potential arise from the bosonic Lagrangian of
SQED with zero mass and a Fayet-Ilyopoulos term
\cite{Edelstein:1993bb}. In this case the ``miracle'' of the
proportionality between the tension and the charge finds its
explanation in the central charge of the supersymmetry algebra
\cite{Witten:1978mh}.\footnote{Even if the context is different, we
just mention that there is another hot line of research in which
non-supersymmetric gauge theories have supersymmetry relics in the
large $N$ limit \cite{Armoni:2004uu}.}


In the present paper we have finaly found a convincing proof that
multi-vortices become bags in the large $n$ limit. This fact is very
general and applies also in the multi-monopole case
\cite{monopolemio}. It will be interesting to see if the bag
mechanism works also in more general theories that contain solitons:
nonabelian vortices \cite{nonabelianvortices}, semi-local strings
\cite{semilocalstring}, Chern-Simon theories \cite{chernsimon} and
noncommutative field theories \cite{noncommutativegeometry}.

\section*{Acknowledgments}
We thank Mads T. Frandsen, Chris Kouvaris, Konstantin Petrov,
Thomas Ryttov and Francesco Sannino for useful discussions. S.B.
wants to thank expecially Roberto Auzzi, Jarah Evslin, Kenichi
Konishi, and Marco Matone for comments and discussions. The work
of S.B. is supported by the Marie Curie Excellence Grant under
contract MEXT-CT-2004-013510 and by the Danish Research Agency.

\end{document}